# Structure and Bonding in Amorphous Iron Carbide Thin Films

Andrej Furlan[1], Ulf Jansson[2], Jun Lu[1], Lars Hultman[1], and Martin Magnuson[1]

[1]*Thin Film Physics Division, Department of Physics, Chemistry, and Biology (IFM), Linköping University, SE-58183 Linköping, Sweden*

[2]*Department of Chemistry – Ångström Laboratory, Uppsala University, Box 538, SE-751 21 Uppsala Sweden*

2015-01-08

## Abstract

We investigate the amorphous structure, chemical bonding, and electrical properties of magnetron sputtered $Fe_{1-x}C_x$ (0.21≤x≤0.72) thin films. X-ray, electron diffraction and transmission electron microscopy show that the $Fe_{1-x}C_x$ films are amorphous nanocomposites, consisting of a two-phase domain structure with Fe-rich carbidic $FeC_y$, and a carbon-rich matrix. Pair distribution function analysis indicates a close-range order similar to those of crystalline $Fe_3C$ carbides in all films with additional graphene-like structures at high carbon content (71.8 at% C). From X-ray photoelectron spectroscopy measurements, we find that the amorphous carbidic phase has a composition of 15-25 at% carbon that slightly increases with total carbon content. X-ray absorption spectra exhibit increasing number of unoccupied *3d* states and decreasing number of C *2p* states as a function of carbon content. These changes signify a systematic redistribution in orbital occupation due to charge-transfer effects at the domain-size dependent carbide/matrix interfaces. Four-point probe resistivity of the $Fe_{1-x}C_x$ films increases exponentially with carbon content from ~200 μΩcm (x=0.21) to ~1200 μΩcm (x=0.72), and is found to depend on the total carbon content rather than the composition of the carbide. Our findings open new possibilities for modifying the resistivity of amorphous thin film coatings based on transition metal carbides by control of amorphous domain structures.

## 1. Introduction

The structure and chemical bonding of binary transition metal carbide compounds are strongly dependent on the transition metal [1,2]. Early transition metals such as Ti, Zr and V form strong metal-carbon bonds in NaCl-type cubic crystals in contrast to the late transition metals such as Fe and Ni that have less strong Me-C bonds [3]. The later transition metals form much more complex structures, where the carbon atoms are found in both trigonal prismatic and octahedral sites in large structural unit cells such as $Fe_3C$ and $Cr_{23}C_6$ [4].

The early transition metals Ti, Nb, Zr and V generally form nanocomposites with nanocrystalline MeC grains with a NaCl-structure (nc-MeC) in an amorphous carbon matrix (a-C) [5]. The size of the carbide grains as well as thickness of the matrix can be controlled by, e.g., changes in carbon content [3,6]. A structural analysis of these materials is also complicated due to interface effects. We have recently demonstrated that charge transfer occurs between the Ti atoms in nc-TiC grains to the carbon atoms in the a-C matrix [5,7]. This charge transfer changes the chemical bonding around the metal atoms in the carbide/carbon interface, and will of course affect the materials properties. In the case of nc-TiC/a-C, the effect of the charge transfer can be seen as an additional bonding state in X-ray photoelectron spectroscopy (XPS) spectra, and an increase of the lattice parameter of the nc-TiC grains [3,5].

In contrast, the later transition metals tend to yield more amorphous films with also non-crystalline carbides. This was recently demonstrated in sputter deposition of $Cr_{1-x}C_x$ films [6,7], which were found to be completely amorphous over a broad composition range with an amorphous chromium carbide phase in an a-C matrix. Amorphous films have also been observed in the Fe-C system, while Ni-C completely crystallize at low carbon content [8]. Bauer-Grosse demonstrated that amorphous $Fe_{1-x}C_x$ films [9,10] consist of two components; a Fe-rich metallic $FeC_y$ phase, and an amorphous carbon-rich phase with less Fe. When the carbon content increases, the amorphous carbon phase accommodates the additional carbon while the composition of the carbidic $FeC_y$ phase remains constant. Also in this case we can expect charge transfer effects at the carbide/matrix interface. This was also demonstrated in an earlier study for sputter-deposited a-CrC/a-C films [7]. However, this effect is expected to decrease for the later transition metals in the *3d* series due to a smaller difference in electronegativity between the metal and carbon.

Consequently, we can have two types of amorphous sputter-deposited Me-C thin films; nc-MeC/a-C with nanocrystalline carbide grains and a completely amorphous nanocomposite a-MeC/a-C. However, the structure and bonding of the completely amorphous nanocomposites are not established. Furthermore, the trend in charge transfer effects for different metals has yet not been investigated. In this work, we are closely investigating the structure and bonding in binary Fe-C films with different carbon contents. By employing high-resolution transmission electron microscopy (HR-TEM) with selected area electron diffraction (SAED) and pair distribution function (PDF) analysis, X-ray photoelectron spectroscopy (XPS), Raman, and soft X-ray absorption spectroscopy (XAS), we obtain information about the amorphous structure and bonding with a special aim to investigate the carbon content in the $FeC_y$ phase and charge transfer effects in Fe-C films in comparison to the Ti-C, Cr-C and Ni-C systems. Here, we use magnetron sputtering to grow the Fe-C films. This technique is useful because it affords films with a wide range of compositions and tuning of structure [11,12].

## 2. Experimental details

*2.1 Synthesis*

$Fe_{1-x}C_x$ films with a high degree of purity, and precisely tuned composition were deposited on single-crystal Si(001) (10x10 mm) substrates by dual dc magnetron sputtering in an ultra high vacuum (UHV) chamber with a base pressure of $10^{-9}$ Pa. The substrates were biased to -50 V, and preheated to 250 ˚C from the back side by a resistive heater built into the substrate holder. The thicknesses of as-deposited coatings were between 170 nm - 470 nm after deposition from a double current regulated 2 inch magnetron set-up in Ar discharge generated at 3.0 mTorr with a gas flow rate of 30 sccm. The magnetrons were directed towards a rotating substrate holder at a distance of 15 cm. As sputtering sources, separate graphite, and Fe targets were used (C 99.999 % pure, and Fe 99.95 % pure). In order to enable the magnetic field from the magnetron to reach the plasma without remaining contained in the ferromagnetic target, two methods were used. The first method, used for the deposition of most of the $Fe_{1-x}C_x$ films, implied that the circular center part of a Fe target was replaced by a graphite plate. In this way a simultaneous sputtering of Fe, and C from the same target was achieved. For the deposition of a pure Fe film as a reference, a Fe target (99.95 % pure) of the thickness of only 1 mm has been used. This thin target enabled the magnetic field of the magnetron to penetrate the target, and come into contact with the plasma. In the second method, the same target, together with a pure graphite target, was used for the deposition of $Fe_{1-x}C_x$ films that served as a comparison to examine the influence of the elemental, and non-elemental targets on $Fe_{1-x}C_x$ structure crystallization. The film composition was controlled by keeping the graphite target at a constant current of 300 mA, and tuning the current on the Fe target. The resulting thicknesses of the as-deposited coatings were 470 nm ($x=0.208$), 440 nm ($x=0.263$), 315 nm ($x=0.445$), 340 nm ($x=0.718$) and 200 nm ($x=1.0$: a-C) as determined by X-ray reflectometry (XRR).

*2.2 Characterization*

The structural properties of the $Fe_{1-x}C_x$ thin films were determined by high-resolution X-ray diffraction (XRD) analysis. In order to avoid diffraction signal from the Si substrate, grazing incidence (GI) XRD measurements were carried out on a PANanalytical EMPYREAN using a Cu $K_\alpha$ radiation source, and parallel beam geometry with a 2° incidence angle to avoid substrate peaks and minimize the influence of texture. Each XRD scan was performed with 0.1° resolution, 0.05° step length with a total of 1800 points for 6 hours. HR-TEM images and selected area electron diffraction (SAED) patterns were obtained with a FEI Tecnai $G^2$ 20 U-Twin 200 kV field emission gun TEM (FEG-TEM). Cross-section samples were mechanically polished, and ion milled to electron transparency by a Gatan Precision Ion Polishing System (PIPS). The PDF analysis was performed by using the selected electron diffraction patterns and analyzing them with RDF-Tools script in Digital Micrograph software [13].

The compositions of the films were determined by X-ray photoelectron spectroscopy (XPS) using a Physical Systems Quantum 2000 spectrometer with monochromatic Al $K_\alpha$ radiation. Depth profiles of the films were acquired by rastered $Ar^+$-ion sputter etching over an area of 2 x 2 $mm^2$ with the ions being accelerated by a potential difference of 4 kV. The high-resolution scans of the selected energy regions were acquired after 45 min, 30 min, and 6 min of $Ar^+$-ion sputter etching with ions being accelerated by the potential difference of 200 V, 500 V or 4 kV, respectively. The XPS analysis area was set to a diameter of 200 μm and the step size to 0.05 eV with a base pressure of $10^{-9}$ Pa during all measurements. The peak fitting was made by Voigt shape functions to account for the energy resolution of the instrument and chemical disorder (Gaussian part) and the lifetime width of the photoionization process (Lorenzian part). In order to correlate the nanostructuring, and $sp^2/sp^3$ ratio of the films to the Fe concentration, Raman scattering spectroscopy was also performed at room temperature in the range 800-1900 cm$^{-1}$ in a back-scattering configuration using UV 325 nm laser excitation.

X-ray absorption (XAS) measurements were performed at the undulator beamline I511-3 at MAX II (MAX-lab Laboratory, Lund, Sweden), comprising a 49-pole undulator, and a modified SX-700 plane grating monochromator [14]. The measurements were performed with a base pressure lower than $6.7*10^{-7}$ Pa. The XAS spectra were measured at 5° grazing incidence angle from the surface plane and a detection angle of 30° from the incident photon direction. All samples were measured in the same geometry with energy resolutions of 0.2, and 0.1 eV at the Fe *2p*, and C *1s* absorption edges, respectively. The XAS spectra were normalized to the step before, and after the absorption edges and corrected for background and self-absorption effects [15] with the program XANDA [16] in Figs. 6, and 7.

Cross-sectional scanning electron microscopy (SEM) images were obtained in a LEO 1550 microscope using accelerating voltages of 15 kV in *in-lens* imaging mode. The obtained images were used for thickness measurements, and structural analysis of the coatings. Sheet resistance measurements were made with a four-point probe "4-dimensions" 280C. For each sample, four readings were made with a different 4-sensor orientation around the center of the sample. As a final value of the electrical resistivity, a mean value over four measurements was made. Each set of measurements on a sample showed similar values indicating negligible influence from surface oxide.

## 3. Results

### 3.1 XRD, TEM and PDF analysis

Figure 1 shows X-ray diffractograms (XRD) of the $Fe_{1-x}C_x$ films for x=0.21, 0.26, 0.45, and 0.72. The absence of sharp diffraction peaks shows that all films are X-ray amorphous. However, a broad structure is observed in the range between 36-50° indicating dominant amorphous character with some partial short-range ordered structure at low carbon content as indicated by the arrow in Fig. 1. To obtain insight into the detailed structure of the $Fe_{1-x}C_x$ films, HR-TEM images were made, as shown in Fig. 2, together with the corresponding SAED patterns and pair-distribution function (PDF) analysis result. Figs. 2(a1-4) show the high-resolution images of the samples with carbon content ranging from 20.8 at% to 71.8 at%. Based on mass contrast, the bright domain boundaries are carbon rich while the dark domains are Fe rich. As can be seen in the HR-TEM images, the domain size is approximately 5 nm in the samples with carbon content less than 30 at%. With increasing carbon content in the samples, the domain size gradually decreases to about 2 nm while the domain boundary regions increases at high carbon content (see Figs. 2a3 and 2a4). The HR-TEM images also show that both the domains and domain boundaries are amorphous, in spite of the appearance of a partly ordered short-range structure in some domains at the low carbon content sample. Non-perfectly ordered 3D crystal structures are observed in the $FeC_y$ carbide phase by the HR-TEM. The amorphous character is exhibited in the corresponding SAED patterns, collected from an area of approximately 250 nm in diameter in Figs. 2(b1-4), as broad rings typical for amorphous structures. The rings become more diffuse with increasing carbon content. The diffuse ring observed at a plane distance spacing of ~2.1±0.4 Å is identical to the broad feature at 2θ = 36-50° as indicated in the diffractograms in Fig. 1. Based on the XRD and the selected area electron diffraction results, we define the $Fe_{1-x}C_x$ carbide phase as amorphous.

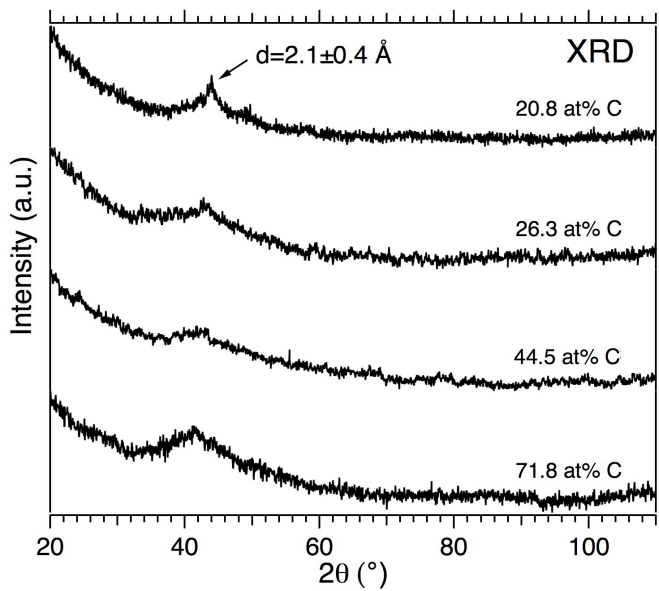

**Fig. 1:** X-ray diffraction (XRD) of the $Fe_{1-x}C_x$ films with C content ranging from 20.8 at.% to 71.8 at.%.

To obtain information on the close-range order, a PDF study was performed and the result is shown as raw data in Figs 2(c1-4). It is likely that an amorphous iron carbide phase is present in the films. If we consider the most well-known iron carbide $Fe_3C$, the Fe-Fe distance in this phase is about 2.5 Å, that agrees well with the strongest peak A in Fig. 2(c1-4). This implies that the nearest neighbor Fe-Fe close-range order is probably similar to that of the crystalline $Fe_3C$ phase. Peak B at ~3.8 Å also agrees with the second coordination shell distance of Fe-Fe in a $Fe_3C$ structure [9]. However, note that the contribution of the C-C and Fe-C bonds in $Fe_3C$ are much weaker compared to the Fe-Fe bonding contribution. Therefore, the signal from expected PDF peaks at ~2.0 Å corresponding to the C-Fe bond and ~ 3.1 Å corresponding to C-C bonds are below the noise level compared to the strong Fe-Fe signal for most of samples because of a much weaker scattering cross section of carbon compared to iron. However, note that for the sample with 71.8 at%

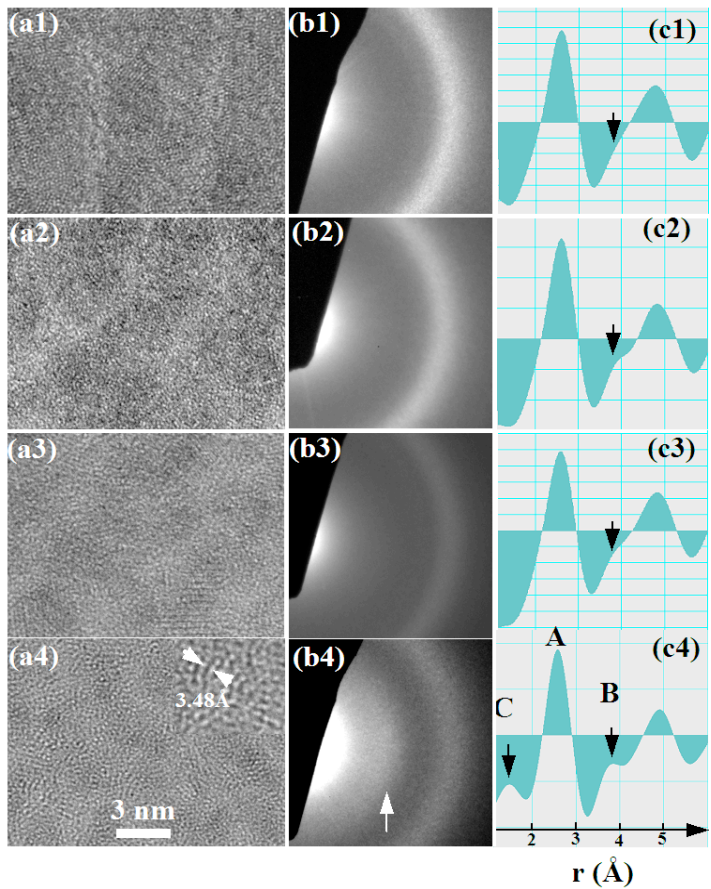

**Fig. 2: Figs. 2:** HR-TEM micrographs (a1-4), SEAD patterns (b1-4) and PDF patterns (c1-4) of $Fe_{1-x}C_x$ samples with carbon content ranging from 20.8 to 71.8 at%.

C, a peak C at ~1.4 Å actually shows up. Peak C can be assigned to the in-plane nearest neighbor C-C bonding distance of graphene-like structures. The second coordination shell distance between C-C in the graphene-like planes is at ~2.48 Å, that overlaps with peak A. A graphene-like structure with an interplanar distance of 3.5 Å is also observed in both HR-TEM and electron diffraction, as shown by the inset in Fig. 2(a4) and white arrow in Fig. 2(b4), respectively.

*3.2 X-ray photoelectron spectroscopy (XPS)*

Figure 3 shows C *1s* core-level XPS spectra of the four $Fe_{1-x}C_x$ films with x=0.21, 0.26, 0.45, and 0.72. As can be seen, at least three peaks are required to deconvolute the spectra. A peak at 283.3 eV can be assigned to C-Fe bonds, while a second peak at 285.0 eV can be assigned to $sp^3$ hybridized carbon (C-C-$sp^3$) [18,19]. Between these two peaks, a third feature is clearly present. The intensity of this peak increases with carbon content and is also shifted from about 284.3 eV for the most Fe-rich film to about 283.9 eV in the most C-rich film. Most likely, several types of carbon is contributing to this feature. Firstly, $sp^2$-hybridized carbon is known to exhibit a peak at about 0.9 eV lower binding energy than $sp^3$-hybridized carbon, i.e. at about 284.1 eV [20,21]. In HR-TEM in Fig. 2, we observed domains with an iron-rich and a carbon-rich phase that is often observed in binary sputter-deposited metal carbide films that are nanocomposites with a carbide phase in an amorphous carbon (a-C) matrix [3,7,8]. Previous C *1s* XPS studies on the a-C phase have shown a mixture of $sp^2$- and $sp^3$-hybridized carbon. Secondly, studies on sputter-deposited Me-C films have also shown an additional Me-C feature at a slightly higher binding energy [3] that can be due to sputter damage of the metal carbide grains. Thirdly, a contribution originates from surface Me atoms in the carbide grains. This is caused by charge transfer effects where charge will be transferred from the metal surface atoms to the more electronegative carbon atoms in the a-C matrix [7]. In nanocomposites with very small grains or domains, the relative amount of surface atoms is large and will show up as a high-energy shoulder on the main C *1s* Me-C peak (denoted Me-C*) [8,9]. For the $Fe_{1-x}C_x$ films, it is impossible to deconvolute the feature at 283.9-284.3 eV into separate C-C ($sp^2$) and Fe-C* peaks. However, a comparison with Ti-C, Cr-C and Ni-C films show that the Me-C* contribution is small compared to the C-C ($sp^2$) peak [5-8]. For this reason, we assign the entire peak at 283.9-284.3 eV to C in a-C although it will give a slight overestimation of the amount of the C-C ($sp^2$) phase compared to the iron carbide ($FeC_y$) phase.

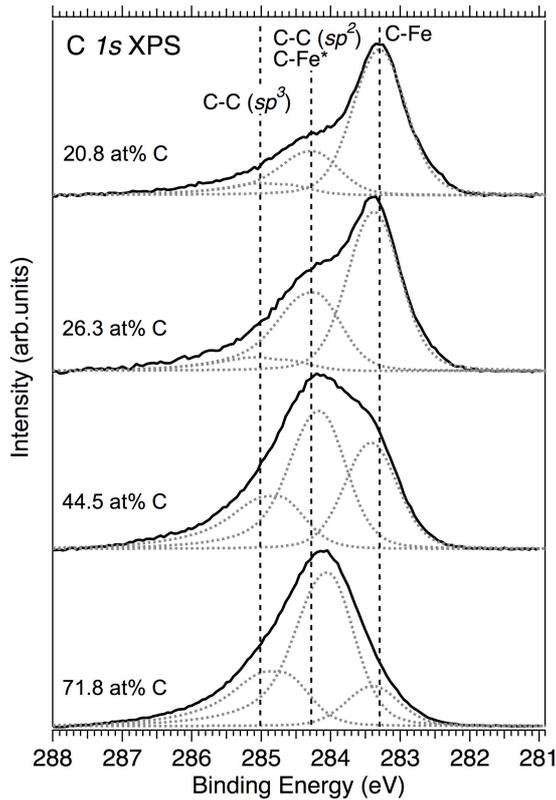

**Fig. 3:** C *1s* XPS spectra of the $Fe_{1-x}C_x$ films with carbon content ranging from 20.8 at.% to 71.8 at.%. The deconvoluted peaks at 283.3, and 285.0 eV, indicated by the dashed vertical lines corresponds to carbidic $FeC_y$ carbon in Fe-C bonds, and free carbon in C-C bonds, respectively. A third structure is identified at 284.3 eV, and can be associated with charge-transfer C-Fe* bonds or C-C in $sp^2$ hybridized bonds.

The XPS data complements the amorphousness of the films indicated by the TEM and XRD studies and show that the films consist of two phases; an amorphous $FeC_y$ carbide phase (a-FeC) dispersed in amorphous carbon (a-C).

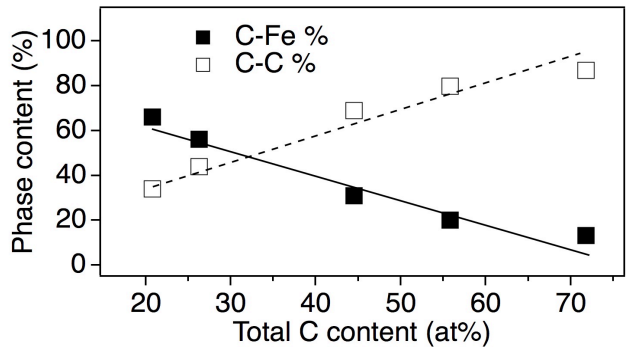

**Fig. 4:** Relative amount of C-Fe, and C-C bonds determined as the proportions of the areas fitted on the C *1s* XPS peak. The linearly fitted lines are guidelines for the eye.

Figure 4 shows the relative amount of the two phases as a function of total carbon content (assuming that the Fe-C* contribution to 283.9-284.3 eV peak can be neglected). As can be seen, the relative amount of the a-C phase increases linearly with the total carbon content. Table I shows the total composition analysis that is valid under the assumption that the total photoelectron cross-section in all samples is constant for carbon. The carbon content in the amorphous $FeC_y$ phase can now be estimated using the data in Fig. 4. The analysis shows that the carbon content of the carbide phase slightly increase with the total carbon content from 15 at% (20.8 at% total), 16 at% (26.3 at % total), 20 at% (44.5 at% total) to 25 at% (71.8 at% total). On average, the carbide domains contain about 20 at% carbon that is close to the carbon content in the crystalline phases $Fe_3C$ ($FeC_{0.33}$) [9], $Fe_4C$ [22] and $Fe_{23}C_6$ ($FeC_{0.26}$) [17]. It is also close to the eutectic point in the binary Fe-C phase diagram of 17.8 at% [23]. However, the carbon content in the carbide phase could be underestimated and therefore represents a lower limit since the contribution of Fe-C* has been neglected in the analysis of the C *1s* spectra.

*3.3 Raman spectroscopy*

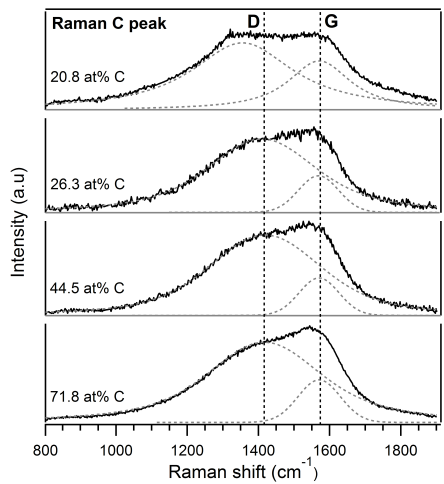

**Fig. 5:** Raman spectra of the carbon peak of the $Fe_{1-x}C_x$ films for x=0.21, 0.26, 0.45, and 0.72. The two vertical dashed lines indicate the disorder (D), and graphite (G) peaks of the fitted peak components [24,25].

Figure 5 shows carbon Raman spectra of the $Fe_{1-x}C_x$ films with *x* values of 0.21, 0.26, 0.45, and 0.72. The two band components in the spectra, the disordered (D), and graphite (G) peaks, were deconvoluted by Voigt shape functions. In pure graphite, the vibrational mode that gives rise to the G-band is known to be due to the relative motion of $sp^2$ hybridized C atoms while the D band is due to the breathing vibrational mode of the six-membered rings [24]. The energies of the D, and the G bands have an almost constant position around 1410 $cm^{-1}$, and 1570 $cm^{-1}$, respectively. The $I_D/I_G$ height ratios of the films are 1.05, 0.87, 0.86, and 0.83, respectively, and are related to the $sp^2/sp^3$ hybridization ratio [25]. The $I_D/I_G$ height ratios are similar for *x*=0.26 and 0.45, but lower for the film with *x*=0.72. On the other hand, for the lowest carbon content (*x*=0.21), the $I_D/I_G$ height ratio observed in the Raman spectra is higher than for the other samples, indicating a lower $sp^2$ fraction. The $I_D/I_G$ ratios approximately correspond to $sp^2$ fractions of 0.60, 0.73, 0.74, and 0.81, respectively. The generally high fraction of $sp^2$ hybridization increases with carbon content as the intensity of the G-peak increases.

*3.4 Fe 2p X-ray absorption spectroscopy*

Figure 6 shows Fe *2p* XAS spectra of the *3d*, and *4s* conduction bands following the Fe $2p_{3/2,1/2} \rightarrow 3d$ dipole transitions of the $Fe_{1-x}C_x$ films with different carbon content. The main peak structures are associated with the $2p_{3/2}$, and the $2p_{1/2}$ core-shell spin-orbit splitting of 13.1 eV. The double-sub-peaks denoted (1)=707.3, (2)=708.7 eV, (3)=720.3 eV, and (4)=721.9 eV correspond to the usual ligand-field splitting (1.4 eV) observed in other transition-metal carbides [7,25,36]. A comparison of the spectral shapes at different carbon contents shows two interesting effects: (i) the total intensity of the spectra increase with carbon content as in the case of Cr *2p* XAS in amorphous $CrC_x$ [7], and Ti *2p* XAS in nc-TiC/a-C nanocomposites that exhibit the same trend [5].

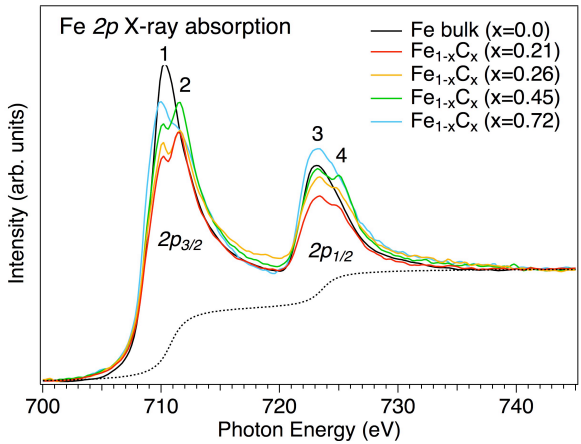

**Fig. 6:** Fe *2p* TFY-XAS spectra of $Fe_{1-x}C_x$ for the different C contents in comparison to bulk Fe (x=0). The dashed curve indicates the integrated background.

The Fe *2p* XAS intensity is proportional to the unoccupied *3d* states, and therefore this trend indicates increased charge-transfer for increasing *x* as the occupied Fe *3d* electron density is reduced around the absorbing Fe atoms for higher carbon concentration. (ii) The XAS spectrum of Fe metal (x=0) has narrower $2p_{3/2}$, and $2p_{1/2}$ absorption peaks, whereas the XAS spectra of the carbon-containing films exhibit small peak structures (1), and (2) at the $2p_{3/2}$ peak. The Fe *2p* XAS spectra mainly represent the iron contribution in the amorphous $FeC_y$ carbide phase. The peak positions of all Fe spectra are essentially the same and independent of *x*, consistent with the constant peak position in Fe *2p* XPS (707.1 eV).

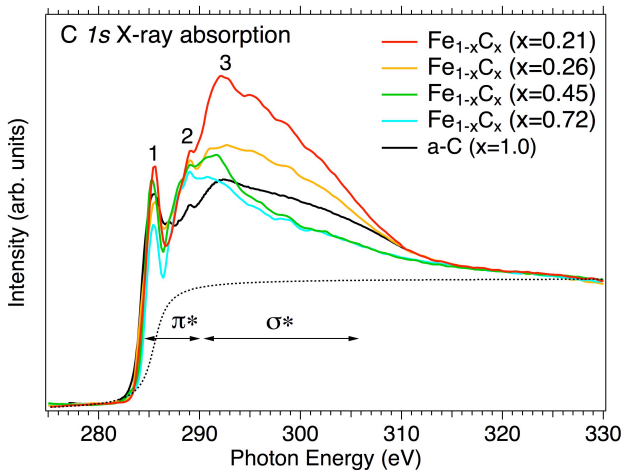

**Fig. 7:** C *1s* TFY-XAS spectra of $Fe_{1-x}C_x$ for different C contents compared to amorphous carbon (a-C, x=1). The dashed curve indicates the integrated background.

A comparison of the integrated peak areas at different carbon contents shows that the $2p_{3/2}/2p_{1/2}$ peak ratio is lowest for *x*=0.72 (1.9), and *x*=0.45 (2.0) in comparison to the lower carbon contents *x*=0.26 (2.2), and 0.21 (2.3), while it is higher for bulk Fe *x*=0 (2.4). A lower $2p_{3/2}/2p_{1/2}$ peak ratio is an indication of higher ionicity (lower conductivity) for the higher carbon contents [26,27]. Note that the $2p_{3/2}/2p_{1/2}$ peak ratio is a result of the ionicity for Fe, in the $FeC_y$ carbide phase, and not for the entire film that is the sum of both the Fe-rich (~$FeC_y$), and carbon-rich domains in the matrix.

When the main part of the film consists of C-rich matrix areas, this phase determines the resistivity as discussed in section 3.6.

*3.5 C 1s X-ray absorption spectroscopy*

Figure 7 shows C *1s* XAS spectra of the $Fe_{1-x}C_x$ films probing the unoccupied C *2p* conduction bands as a superposition of the $FeC_y$ carbide phase, and the changes in the C matrix phase. The intensity of the C *1s* XAS spectra generally follow an opposite trend in comparison to the Fe *2p* XAS spectra, and indicate increasing charge-transfer from Fe to C for increasing C-content (more occupied C states for larger *x*). The first peak structure (1) at ~285 eV is associated with empty π* states, and the higher states above 290 eV are due to unoccupied σ* states. The empty π* orbitals consists of the sum of two contributions in $Fe_{1-x}C_x$: (i) $sp^2$ (C=C), and $sp^1$ hybridized C states in the amorphous carbon phase, and (ii) C *2p* - Fe *3d* hybridized states in the amorphous iron carbide phase. Peak (2) at 288.3 eV is also due to C *2p* - Fe *3d* hybridization with addition of the superimposed carbon phase [7], and possible C-O bond contributions. The energy region above 290 eV is known to originate from $sp^3$ hybridized (C-C) σ* resonances, where peak (3) at 295 eV forms a broad shape resonance with multielectron excitations towards higher energies [7]. The broad structure (3) shows highest intensity for *x*=0.21 that originates from $sp^3$ hybridized σ* states, consistent with the Raman observations.

By calculating the integrated π*/[π*+σ*] intensity ratio, an estimation of the relative amount of π* ($sp^2$, $sp^1$ hybridization) content in the samples was made following the procedure in Ref. [28], and presented in Table I. The integrated π*/[π*+σ*] intensity ratio was calculated by first fitting an integrated step-edge background. Thereafter, Gaussian functions were fitted to the π*+σ* peaks, following the procedure in Refs. [37, 38] and presented in the last column of Table I. We assumed that π* peaks occur below and σ* peaks above 290 eV as indicated in Fig. 7. This analysis method gives an estimation of the relative amount of π* ($sp^2$, $sp^1$) hybridization content in the samples. The fraction of $sp^2$ is smallest for *x*=0.21 and highest for *x*=0.72, following a similar trend as the XPS and the Raman results. However, the $sp^2$ fraction obtained from XPS is overestimated at low carbon content as the at% C in the $FeC_y$ phase is underestimated.

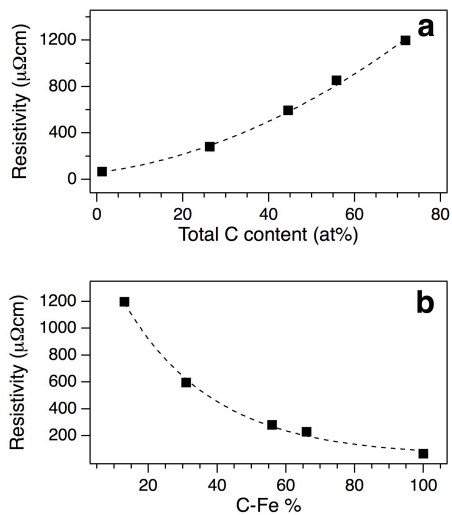

**Fig. 8:** a) Resistivity of the $Fe_{1-x}C_x$ films depending on the C content as determined from sheet resistance measurements by a four-point probe, and film thickness determined by SEM. b) The resistivity is plotted as a function of the relative amount of C bound as C-Fe. The dashed least-square fitted curves are guides for the eye. As a reference, the resistivity of an almost pure Fe thin film sample of thickness comparable to the other samples is shown for comparison.

*3.6 Resistivity measurements*

Figure 8a shows the electrical resistivity dependence of the C content for the investigated $Fe_{1-x}C_x$ films. For the film with highest C content, the resistivity is as high as 1200 μΩcm. With decreasing amount of C content, the resistivity decreases exponentially ending at a minimum value of ~70 μΩcm for a nearly purely metallic α-Fe film containing 1.2 at% of C. Compared to the electrical resistivity of 9.98 μΩcm for pure bulk iron [21], this value is almost an order of magnitude higher. However, a higher value than for bulk Fe is expected as it is caused by, not only the small carbon content, but also the structure of this film that consists of amorphous domains with relatively small size. It is also known that pure metallic thin films display higher electrical resistivity compared to the bulk metals [29, 30]. The increase of electrical resistivity with the increasing C content in the films is correlated to the increased amount of C-C bonding and the substitutional incorporation of C atoms in the Fe lattice sites forming the $FeC_y$ carbide phase. The C-rich matrix phase is known to be a very poor conductor compared to pure Fe metal, while for the bulk $FeC_y$ carbide phase, it has been suggested [31] that the incorporation of C in Fe increases electrical resistivity by around 6 μΩcm/at%. Figure 8b shows the change of electrical resistivity in relation to the proportion of the $FeC_y$ carbide phase. As shown, the resistivity is lowest when there are many C-Fe bonds. In order to decrease the resistivity below 400 μΩcm, the proportion of the C-rich phase should be less than 30at%. Although the resistivity of the pure amorphous $FeC_y$ carbide phase component is difficult to determine, we estimate that the resistivity of that phase is higher than the crystalline $Fe_3C$ phase, or other Fe-C crystals, for a given C concentration.

**Table I:** Composition of the amorphous $Fe_{1-x}C_x$ films for $x$=0.21, 0.26, 0.45, and 0.72. The C-C ($sp^3$), C-C ($sp^2$) and C-Fe chemical bond contributions and $sp^2$ fractions were determined by integrating the areas under the corresponding peak structures in the C $1s$ XPS spectra. The Raman $sp^2$-fraction was estimated from Ref. 23.

| Composition | $Fe_{0.79}C_{0.21}$ | $Fe_{0.74}C_{0.26}$ | $Fe_{0.55}C_{0.45}$ | $Fe_{0.28}C_{0.72}$ |
|---|---|---|---|---|
| at% C in $FeC_y$ phase | 15 | 16 | 20 | 25 |
| XPS $sp^2$ fraction | 0.75 | 0.82 | 0.67 | 0.69 |
| Raman $sp^2$ fraction | 0.60 | 0.73 | 0.74 | 0.80 |
| C $1s$ XAS: $\pi^*/[\pi^*+\sigma^*]$ | 0.52 | 0.51 | 0.66 | 0.79 |

## 4. Discussion

The combined observations in this study show that the $Fe_{1-x}C_x$ films are X-ray amorphous nanocomposites, and consist of two amorphous phases; (i) an amorphous carbide phase, and (ii) an amorphous carbon matrix, consistent with other observations [32-35]. The XPS analysis confirms the two-phase observations in HR-TEM with an iron-rich carbide phase and an amorphous carbon-rich phase. As determined from the XPS results shown in Fig. 3 and 4, the C-C and C-Fe binding energies are essentially the same for all compositions. Generally, the 1.7 eV difference in binding energy of the C-Fe peak in comparison to the C-C peak signifies electronic charge-transfer from Fe to C. This chemical shift in XPS is smaller than in the case of Ti to C [8,36] due to the smaller difference in electronegativity for Fe and C.

Contrary to previous observations in a smaller composition range [10,11], we find that the carbon content of the carbidic $FeC_y$ phase is not constant, but slightly increases from 15 to 25 at% with the total carbon content. The 20 at% average carbon content agree well with the composition of the crystalline iron-carbon carbides $Fe_3C$ (y=0.33), and $Fe_{23}C_6$ (y=0.26) but the structure of the investigated samples is amorphous. However, the C-Fe* contribution is difficult to separate from other contributions in the XPS spectra as it is an overlapping part of the third peak between the C-C and C-Fe peaks at 283.9-284.3 eV that should be associated with charge-transfer effects from the carbide/matrix interface between the amorphous domains. This results in an overestimation of the amount of C-C in the $sp^2$-phase and the average of 20 at% carbon content in the carbide phase thus represents a lower limit due to the neglection of the Fe-C* contribution.

For the $x$=0.21 sample, the broad diffraction ring at 2.1±0.4 Å in the SAED corresponds to the broad XRD structure observed around 2θ = 36-50° in Fig. 1 that represent short-range order plane distances. Note that the bond distance and plane distance are different, where the latter corresponds to the peaks in XRD, and SAED. The PDF peak at 2.5 Å agrees with the shortest Fe-Fe bond distance in a $Fe_3C$-like structure. The PDF peak position is independent on carbon content in the samples, indicating a constant composition in the carbide structure. According to our composition analysis, the amorphous carbide $FeC_y$ phase can be either $Fe_3C$ or $Fe_{23}C_6$ like. $Fe_3C$ is a more common phase and can relatively easy be formed. On the other hand, $Fe_{23}C_6$ has a much more complex structure that can only be formed at high temperature. Thus, the amorphous $FeC_y$ carbide phase shows a short-range order similar to that of a $Fe_3C$ structure, but further studies are required. As observed by the HR-TEM, the size of the amorphous carbidic $FeC_y$ domains decreases exponentially from 5±1 nm for $x$=0.21 and 0.26 to, 4±1 nm for $x$=0.45, and 2±1 nm for $x$=0.72. The same type of trend was observed for nanocrystalline TiC grains in nc-TiC/a-C nanocomposites [7]. For nanocrystallites such as TiC, the decrease of grain size is generally observed in sputter deposited carbide films and was determined to be due to agglomeration of C atoms at grain boundaries that disrupts further grain growth of crystallites [3]. It is likely that the same type of mechanism is responsible for the size-dependent amorphous domains observed in amorphous nanocomposites.

The high amount of estimated $sp^2$ fraction determined from the peak areas in XPS is confirmed by the Raman data for the two samples that contain the highest carbon content ($x$=0.45, 0.72). For the low carbon content samples ($x$=0.21, 0.26) the $sp^2$ fraction estimated from the XPS data is higher than those estimated from Raman. The difference is related to the curve-fitting procedure or the C-Fe* interface contribution that is not taken into account in the XPS analysis or due to the higher degree of uncertainly in the determination of the peak height and area in the curve fitting procedure for the lower carbon contents. For high carbon contents, the Raman peaks are more resolved and the curve fitting more reliable than for low carbon contents. Another uncertainty in the amount of $sp^2$ determination in Raman is the conversion from peak height ratios of the amplitudes or integrated areas to the $sp^2$ fraction [23]. The $\pi^*/[\pi^*+\sigma^*]$ intensity ratio determined from the C $1s$ XAS spectra is proportional to the $sp^1+sp^2$-fraction and is highest for the highest carbon content of $x$=0.72 in agreement with the Raman estimations. For amorphous $Cr_{1-x}C_x$ [7], at high carbon contents, graphene-like structures were also observed and this is also the case here for the amorphous $Fe_{1-x}C_x$ film at $x$=0.72.

The Fe $2p$, and C $1s$ XAS spectra of $Fe_{1-x}C_x$ (Fig. 6, and 7) exhibit the same general intensity trend as in nanocrystalline TiC embedded in an amorphous C-matrix [5], with large changes in the intensity of the XAS spectra that were attributed to charge transfer effects across the carbide/matrix interface. The intensity of the bulk-sensitive Fe $2p$ XAS spectra generally increase as the domain size decreases that shows that the number of empty Fe $2p$ states increases. An increasing number of unoccupied states in Fe, and decreasing number of unoccupied C states for increasing C-content is observed due to the difference in size of the amorphous $FeC_y$ domains. This reflects a change in the density of unoccupied states, and is attributed to a reduced Fe $3d$ electron density at higher carbon contents. In $Fe_{1-x}C_x$, we expect that charge-transfer occur within the domains of the $FeC_y$ carbide phase, but more significant at the interface to the surrounding amorphous C-phase or between domains, that depends on the domain size.

The average $FeC_y$ carbide composition (15-25 at%) is close to the rhombohedral $Fe_3C$ phase [21] and the more complicated $Fe_{23}C_6$ structure [27] that is a cubic phase that is partly based on prismatic coordination and is known to play an essential role as short-range building blocks in metal glass formation [4,11]. An amorphous mixture of such coordination polyhedra likely occur in the $FeC_y$ domains, where the distribution depends on the total carbon content in the samples. The increasing contribution from the interface bonding as the domain size decreases and the carbon-rich matrix increases with increasing total carbon content affect physical properties such as resistivity, and is therefore essential to take into account in the design of new materials. This is subject to further spectroscopic, and theoretical studies including the effect of magnetic properties.

## 5. Conclusions

We investigated the characteristics of chemical bonds and structure in magnetron sputtered X-ray amorphous nanocomposites of iron-carbide thin films in a broad composition range by X-ray diffraction and high-resolution transmission electron microscopy together with selected area electron diffraction and pair-distribution function analysis, X-ray photoelectron spectroscopy, Raman spectroscopy, and soft X-ray absorption spectroscopy in a broad composition range. From the transmission electron microscopy with selected area electron diffraction and pair-distribution function analysis, and X-ray photoelectron spectroscopy, we observe two types of amorphous domain structures in the $Fe_{1-x}C_x$ films ($0.21 \leq x \leq 0.72$); i) an amorphous $FeC_y$ carbidic structure with a slowly varying carbon content of 15-25%, and ii) an amorphous carbon-rich matrix that accommodates most of the surplus carbon. The X-ray photoelectron, and Raman analysis mainly shows the presence of both Fe-C, and C-C bonds originating from the iron-rich carbidic phase, and the amorphous carbon-rich phase with a high $sp^2$ fraction in carbon. The structure of the $Fe_yC$ carbide has a short-range order similar to that of a $Fe_3C$ structure. Charge-transfer effects at the carbide/matrix interface states represent a third type of phase that increases with carbon content as observed by the changes in orbital occupation in Fe *2p* X-ray absorption spectra. The exponential increase of electrical resistivity with increasing carbon content is correlated with the size of the amorphous iron-rich carbide domains which size decreases while the carbon-rich domains exponentially increases with carbon content and non-linearly affect the resistivity of the films.

**Acknowledgements**

We would like to thank the staff at the MAX IV Laboratory for experimental support, and Jill Sundberg, UU, for help with the Raman measurements. The work was supported by the Swedish Research Council (VR) by project, and Linnaeus grants. M. M., U. J. and J. L. also acknowledge the Swedish Foundation for Strategic Research Synergy Project *FUNCASE*.